\documentstyle[twoside,fleqn,espcrc2]{article}

\title{Origins and renormalization of the superparticle spectrum}

\author{Nir Polonsky\address{Department of Physics and Astronomy,
Rutgers University, Piscataway, NJ 08855-0849, USA}
        \thanks{Work supported by NSF grant No.~PHY-94-23002.}
\thanks{Talk presented at the Fifth International conference on
Supersymmetries in Physics, Philadelphia, May 1997.}
}
       
\begin{document}

\begin{abstract}
The importance of Yukawa contributions to the renormalization of the spectrum 
in non-minimal supersymmetric models is illustrated
in the cases of explicit lepton number violation
(leading to the possibility of singly produced sneutrinos at LEP energies),
an intermediate scale singlet neutrino and negative mass squared parameters 
(possibly modifying fine-tuning considerations),
and a grand-unified sector. The relevance of model-dependent 
renormalization to the supersymmetric flavor problem is emphasized. 
Sources of the flavor problem, some of which
are newly identified, as well as possible solutions, are discussed  
and classified.  It is then shown that gravitational interactions
could lead (via a quadratically divergent singlet) to simple
realizations of some of the low-energy frameworks that attempt to
resolve  the flavor problem. 

\vspace{0.3cm}

\begin{center}
{\large Report No. RU-97-48}
\end{center}

\end{abstract}

\maketitle

\section{Introduction}
Low-energy supersymmery offers an attractive, consistent, and well
motivated extension of the Standard Model (SM).
Whether supersymmetry is indeed realized at the weak scale will be determined
by experiment during  the next decade or so. In particular,
the characterization of the spectrum of the superpartners of
the ordinary fermions (sfermions) and of the gauge and Higgs bosons 
(gauginos and Higgsinos) has to await their  discovery.
(Some hopes that  precision measurements would give clear indirect indications
of certain light sparticles, $e.g.$, from $Z \rightarrow
b\bar{b}$,  proved premature~\cite{H,DP}.)

Once the various masses and mixing angles
are measured, their correlations would allow
one to disentangle and determine the spectrum parameters and 
to learn about its high-energy origins.
The latter  task is complicated by the sensitivity of the spectrum evolution
from high to low energies
to new interactions and to new sectors in the theory
(which couple to the ordinary particles). Such sensitivities, 
on the other hand,
offer unique opportunities to discover new interactions and sectors.

In the absence of solid experimental evidence
we are left, at present, with the following missions:
$(i)$ Surveying and understanding
the available parameter space and the possible spectrum patterns.
$(ii)$ Employing theoretical criteria 
in order to identify the more interesting and motivated possibilities.
The optimal search strategies that are and will be employed
are highly dependent on the specific spectrum pattern one is searching for. 
They are particularly dependent on the identity of the lightest
(and the next to lightest) supersymmetric particle ((N)LSP)
and on the mass hierarchy, which determine the collider signature
(missing energy, hard photons, hadronic activity, etc.) and
the various decay modes, respectively.
Hence, the identification and classification of the possible and the 
more motivated patterns is crucial for a fruitful 
experimental search and analysis (if indeed supersymmetry is realized
in nature). Furthermore, creating a map
between different spectra and the different scenarios of
supersymmetry breaking which they parameterize, may enable us
to have an indirect  glimpse at the supersymmetry breaking
(SSB) sector and at any other coupled sectors in the foreseeable future.

The issues we alluded to above
are vast and are intensively (but far from
completely) studied. Here, we would like to comment on both the
possible origins and the renormalization of the spectrum.
In particular, we will stress that neither
the model dependent spectrum renormalization (or evolution)
nor the ``spectrum pattern - SSB sector'' map are unique. We will also
comment on the supersymmetric flavor problem and on its
role as a selection criterion for models.

We note in passing that one could also choose to study
the dimensionless couplings in the Lagrangian whose form is
dictated by supersymmetry. Any small deviations  from the
supersymmetric identities ({\em e.g.}, between the gauge and gaugino 
couplings to matter and between the gauge and quartic couplings)
could also shed light on heavy and extended sectors.
These issues and the related superoblique parameters
were discussed in these proceedings~\cite{JLF} and in recent
publications~\cite{so,so2}, and will not be explored here.

\section{Minimal supergravity: A reference framework}

Before discussing any complexities, it is instructive to review the
simplistic but phenomenologically consistent and highly predictive
framework of minimal supergravity.
The minimal supergravity framework will also serve us as 
a convenient reference point. (See Ref.~\cite{Nath}
for review and references to the many works that have established
this framework in the last two decades.)

The minimal supergravity framework specifies all three 
functions that define the effective $N = 1$ supersymmetric theory:
$W$, $K$, and $f_{\alpha\beta}$. 
The superpotential 
$W = W_{\mbox{\tiny hidden}} \oplus W_{\mbox{\tiny observable}}$ 
is a direct sum of SSB sector superpotential 
$W_{\mbox{\tiny hidden}}$ 
and of the superpotential corresponding
to the (two-Higgs doublet) SM (suppressing flavor indices
and using self explanatory notation), 
$W_{\mbox{\tiny observable}} = h_{E}H_{1}LE^{c} + h_{D}H_{1}QD^{c}
- h_{U}H_{2}QU^{c} - \mu H_{1}H_{2}$. (Note that the limit~\cite{FPT}
$\mu \rightarrow 0$ was recently excluded 
by searching for an excess of $W^{+}W^{-}$-like events in the LEP2
$\sqrt{s} = 161$ GeV data~\cite{opal}.)

The most simple example of the hidden SSB sector superpotential
is that of Polonyi~\cite{polonyi}, $W_{\mbox{\tiny hidden}} = m^{2}(Z + \beta)$
where $m^{2} \sim {\cal{O}}(m_{\mbox{\tiny{weak}}}M_{P})$ 
determines the scale of supersymmetry breaking and
$\beta$ is a dimensionful constant of the order of the Planck mass $M_{P}$.
($\beta$ is tuned to cancel the cosmological constant).
More generally, one requires $\langle Z \rangle \sim M_{P}$,
$W_{\mbox{\tiny hidden}} \sim m_{\mbox{\tiny{weak}}}M_{P}^{2}$, and
$(\partial W / \partial Z) \sim   m_{\mbox{\tiny{weak}}}M_{P}$.
The gravitino mass $m_{3/2} \sim W_{\mbox{\tiny hidden}}/M_{P}^{2} 
\sim M_{SUSY}^{2}/M_{P}$ is 
$m_{3/2} = {\cal{O}}(m_{\mbox{\tiny{weak}}})$ in this case.
($M_{SUSY}$ is the scale of supersymmetry breaking in the hidden sector
and $M_{SUSY} = m$ in the Polonyi model.)

The Kahler function $K= \Lambda_{ab}\Phi^{a}\Phi^{b\dagger}$
is minimal with $\Lambda_{ab} = \delta_{ab}$ and does not mix the
hidden and observable sectors. The above assumptions 
regarding $W$ and $K$ are sufficient
to ensure that SSB is mediated to the observable sector
gravitationally (and softly) 
and that the (tree-level) boundary conditions for the scalar
potential parameters  are universal (and proportional), {\em i.e.},
at Planckian scales $V_{\mbox{\tiny{SSB}}} 
= m_{0}^{2}\sum_{i}|\phi_{i}|^{2} + (B\mu H_{1}H_{2}
+ A_{0}h_{ijk}\phi^{i}\phi^{j}\phi^{k}$ + h.c.) with 
the soft SSB parameters $m_{0} \sim A \sim B \sim
m_{3/2} \sim m_{\mbox{\tiny{weak}}}$. We also used $\Lambda_{ab}
= \delta_{ab}$ in the summation (otherwise the first term is not universal).
In addition, if assuming
gauge coupling unification (or universal gauge kinetic functions
$f_{\alpha\beta}^{i} = \delta_{\alpha\beta}/g^{2}[1 +  
{\cal{O}}(M_{P}^{-1})]$ at Planckian scales)
then the gauginos also have a common mass at the high-energy boundary
which is parameterized by $M_{1/2} \sim m_{3/2}$.
(It is generated, in principle, from the field-dependent 
terms in $f_{\alpha\beta}$.)
Various mechanisms allow one to also realize
$\mu = {\cal{O}}(m_{3/2})$, so that all dimensionful parameters
in the potential of the observable  sector
are of the order of the weak scale.

Neglecting Yukawa interactions, the renormalized scalar spectrum
simply reads $m_{i}^{2}(Q) = m_{0}^{2} + a_{i}(Q)M_{1/2}^{2}$, where the
momentum-dependent positive-definite
coefficients $a_{i}(Q)$ are charge dependent but flavor blind.
Also, the trilinear terms in the scalar potential are
diagonalized simultaneously with the SM Yukawa matrices.
Both results lead to suppression of new contributions
to flavor changing neutral currents (FCNC).
Such contributions could be unacceptably large for an arbitrary
spectrum (the supersymmetric flavor  problem)
and their successful suppression is crucial for the validity of any model.

The missing ingredient which is required in order to reproduce the
Standard Model (SM) Lagrangian properly is the negative squared mass
in the Higgs potential. Indeed, the $m_{H_{2}}^{2}$ parameter is
differentiated from all other squared mass parameters once we include
the Yukawa interactions. For simplicity, and without loss of
generality, let us assume that only the $t$-quark Yukawa coupling
$h_{t}$ is relevant.  (More generally, the $b$-quark and $\tau$-lepton
couplings may not be negligible.) Then, the well-known one-loop
evolution of $m_{H_{2}}^{2}$ (and of the coupled parameters
$m_{U_{3}}^{2}$ and $m_{Q_{3}}^{2}$) with respect to the logarithm of
the momentum is given by
\begin{equation}
\frac{\partial m_{H_{2}}^{2}}{\partial \ln Q} = \frac{1}{8\pi^{2}}
(3h_{t}^{2}\Sigma_{m^{2}} - 3g_{2}^{2}M_{2}^{2} - g_{1}^{2}M_{1}^{2}),
\label{rge1a}
\end{equation}
and
\begin{equation}
\frac{\partial m_{U_{3}}^{2}}{\partial \ln Q} = \frac{1}{8\pi^{2}}
(2h_{t}^{2}\Sigma_{m^{2}} - \frac{16}{3}g_{3}^{2}M_{3}^{2} -
\frac{16}{9}g_{1}^{2}M_{1}^{2}),
\label{rge1b}
\end{equation}
\begin{eqnarray}
\frac{\partial m_{Q_{3}}^{2}}{\partial \ln Q} = \frac{1}{8\pi^{2}}
(h_{t}^{2}\Sigma_{m^{2}} - \frac{16}{3}g_{3}^{2}M_{3}^{2} -
3g_{2}^{2}M_{2}^{2} && \nonumber \\ - \frac{1}{9}g_{1}^{2}M_{1}^{2}),
\label{rge1c} &&
\end{eqnarray}
where $\Sigma_{m^{2}} = [m_{H_{2}}^{2} + m_{Q_{3}}^{2} +
m_{U_{3}}^{2}+ A_{t}^{2}]$, and we denote the SM SU(3), SU(2) and U(1)
gauge coupling and gaugino mass by $g_{3,2,1}$ and $M_{3,2,1}$,
respectively.  Given the heavy $t$-quark one has $h_{t} \sim 1 \sim
g_{3}$.  (In fact, typically $h_{t} > g_{3}$ at high energies.)  
While QCD loops still dominate the evolution of the stop squared masses
$m_{Q_{3}}^{2}$ and $m_{U_{3}}^{2}$, Yukawa loops dominate the
evolution of $m_{H_{2}}^{2}.$ On the one hand, the stop squared masses
and $\Sigma_{m^{2}}$ increase with the decreasing scale. On the other
hand, the greater they increase the more the Higgs squared mass decreases
with scale, and it is rendered negative at the weak scale.  The $B\mu$
Higgs doublet mixing ensures that both Higgs doublets have
non-vanishing expectation values.
This is a simplistic description of the
well known mechanism of radiative electroweak symmetry breaking.  In
fact, the sizeable $h_{t}$ typically renders the Higgs squared mass
too negative and some (fine?) tuning (usually of $\mu$) is required in
order to extract the precisely known electroweak scale correctly.

Before considering extended and more complicated scenarios, let
us recall the main features of the model:
\begin{itemize}
\item The only interactions between the SSB and observable
sectors are gravitational, {\em i.e.}, the SSB sector is hidden.
 (It is simple, minimal, attractive, but not unique.)
\item
Large Yukawa couplings and Yukawa loops generate a negative
mass squared. (It is a desirable dynamical mechanism, it predicts 
that the $t$-quark  is sufficiently heavy, and it is very attractive.)
\item
One obtains scalar mass universality at the Planckian boundary
at the price  of strong assumptions regarding
the Kahler potential (a desirable result but  unattractive assumptions).
\end{itemize}
Below, we will examine similar effects of 
(strong) Yukawa interactions due to new interactions and new sectors
and their consequences,
look at the issue of universality and the corresponding flavor
problem, and point out new mechanisms which were recently proposed  
for gravitational mediation of SSB~\cite{NP} which are significantly
different from the supergravity framework described above.

\section{Imprints of Yukawa interactions on the scalar spectrum}

Similar dynamics to those leading to the successful
radiative electroweak symmetry breaking appear in many other cases.
The imprints of such dynamics on the scalar potential and spectrum
depend on the strength of the couplings, the corresponding
quantum numbers, and the relevant energy regime.
Once we have understood the importance of loops $\propto h_{t}^{2}$,
it is straightforward to observe and understand
the imprints of extended Yukawa interactions.
We will discuss three examples of such extensions.

\subsection*{New Interactions}

It is well known that 
$W_{\mbox{\tiny observable}}$ could be extended by either lepton 
or baryon number violating Yukawa terms, which are allowed by all 
gauge symmetries, 
$\lambda LLE^{c} + \lambda^{\prime} LQD^{c}$ and
$\lambda^{\prime\prime} U^{c}D^{c}D^{c}$, respectively
~\cite{b}. 
Phenomenological constraints allow for some of the couplings to
be ${\cal{O}}(1)$. If such couplings exist then, {\em e.g.},
\begin{equation}
\frac{\partial m_{L}^{2}}{\partial \ln Q} = \frac{1}{8\pi^{2}}
(3\lambda^{\prime 2}\Sigma_{m^{2}}^{L}
- 3g_{2}^{2}M_{2}^{2} - g_{1}^{2}M_{1}^{2}),
\label{rge2}
\end{equation}
where $\Sigma_{m^{2}}^{L} =[m_{L}^{2} + m_{Q}^{2} + m_{U}^{2}+ 
A_{\lambda^{\prime}}^{2}]$. Eq.~(\ref{rge2}) is obvious
from eq.~(\ref{rge1a}), and all other relevant equations are modified
in a similar fashion. Hence, the slepton doublet
tends to be light (for a large coupling) due to wavefunction
renormalization $\propto \lambda^{\prime 2} m_{\mbox{\tiny{squark}}}^{2}$. 
The successful radiative symmetry breaking mechanism 
implies that if $\lambda^{\prime} \sim {\cal{O}}(h_{t})$
then the relevant slepton doublet is light
(most probably $L_{\tau}$, which we will also denote as $ L_{3}$).
(Note that the $\lambda^{\prime}$ and the
ordinary Yukawa couplings could have simultaneous 
quasi-fixed points~\cite{bbpw}.)
These observations are independent of any supergravity assumptions.

One finds in this case a highly interesting situation: A  strongly interacting
light slepton doublet. 
Together with Erler and Feng we recently
explored such a scenario in some detail~\cite{EFP}. Specifically, we considered
the operators $\lambda_{131}L_{1}L_{3}E_{1}^{c} 
+ \lambda_{333}^{\prime} L_{3}Q_{3}D^{c}_{3}$.
One has the constraints\footnote{We implicitly assume here the basis in which
the sneutrino interaction with the down-type quarks is flavor
diagonal. See Ref.~\cite{EFP} for a general discussion.} 
$\lambda_{131} < 0.1[m_{\mbox{\tiny{slepton}}}/100 {\mbox{ GeV}}]$ 
(from charged current universality) \cite{bgh},
$\lambda_{333}^{\prime} < 0.96[m_{\mbox{\tiny{squark}}}/300 {\mbox{ GeV}}]$ 
(from $B \rightarrow \tau\bar{\nu}X$) \cite{EFP},
and $\lambda_{131}\lambda_{333}^{\prime} < 0.075
[m_{\mbox{\tiny{slepton}}}/100 {\mbox{ GeV}}]^{2}$ 
(from $B \rightarrow e\bar{\nu}$) \cite{EFP}.
The constraints allow for a large $\lambda^{\prime}$,
consequently leading to the possibility of 
light (left-handed) stau $\tilde{\tau}_{L}$
and sneutrino $\tilde{\nu}_{\tau}$.
Their masses are related by SU(2) invariance but 
are slightly split by the electroweak $D$-terms 
$m_{\tilde{\tau}_{L}}^{2} - m_{\tilde{\nu}_{\tau}}^{2} \sim 
-0.77M_{Z}^{2}\cos 2\beta > 0$ (here $\tan\beta =
\langle H_{2} \rangle /\langle H_{1} \rangle $).
Hence, the sneutrino may very well be the LSP (NLSP if the gravitino
is the LSP) and 
$\tilde{\tau}_{L}$ the NLSP, which we will assume in this example.

It is interesting 
to note that the CP even and odd components of the sneutrino
are also slightly split in mass in this case~\cite{EFP}.
This is due to a $b$-quark loop $\propto 
\lambda_{333}^{\prime 2}  m_{b}^{2}$ which 
contributes to the $\tilde{\nu}\tilde{\nu}$ mass term
but not to the $\tilde{\nu}\tilde{\nu}^{*}$ term. 
(The analogue $b-\tilde{b}$ loop generates a neutrino mass,
but it depends also on left-right mixing parameters.)
The split effect was found to be negligible in our case.
(For other examples, see Ref.~\cite{HGG}.)

The large coupling and the light $\tilde{\tau}_{L}$ allow for
$t\rightarrow b\tilde{\tau}_{L} \rightarrow bbc$.  The corresponding
branching ratio (and hence, mass and coupling) can be constrained most
efficiently by the reconstruction of $M_{W}$ in lepton + jets double
$b$-tagged top events at the Tevatron (from the untagged jets).  The
analysis, however, is currently constrained by the limited statistics.
Turning to the sneutrino, 
$e^{+}e^{-} \rightarrow \tilde{\nu}_{\tau} \rightarrow bb$, a fit
to electroweak data assuming a light sneutrino can improve the SM
fit. It suggests $m_{\tilde{\nu}_{\tau}} = 91.79 \pm 0.54$ GeV with
$\lambda_{131} = {\cal{O}}(0.01)$ and $\lambda_{333}^{\prime} =
{\cal{O}}(0.5)$ (from the favorable contributions to the $b$ branching
ratio $R_{b}$ and forward-backward asymmetry $A_{FB}^{0}(b)$).  These
results are highly non-trivial and should be compared to the typically
poor fits in the lepton number conserving cases~\cite{DP}.  Of course,
this does not provide any evidence for $m_{\tilde{\nu}_{\tau}} \simeq
M_{Z}$, but only indicates a window which is slightly preferred by
current data.  More relevant are the search opportunities at LEP2.
The singly produced sneutrino may have a wide resonance
$\sim 6 {\mbox{ GeV}} \times \lambda_{333}^{\prime 2}[m_{\tilde{\nu}_{\tau}}
/ 100 {\mbox{ GeV}}]$.
When the effect of radiative returns ({\em i.e.}, initial state
hard photon radiation) to the sneutrino threshold is also included, 
one finds that a low-luminosity scan with only a few steps 
can either discover the sneutrino at LEP2 energies 
(up to the kinematic limit) or
significantly improve the constraints on the couplings.  In certain
regions of the parameter space the anticipated $\sqrt{s} \sim 180 -
190$ GeV runs may suffice for discovery of the sneutrino. (For
detailed examples and discussion, see Ref.~\cite{EFP}.)

It should be noted that the sneutrino tail would enable
one to explore sneutrinos slightly above the kinematic limit.
Also, the $e^{+}e^{-} \rightarrow \tilde{\nu}_{\tau} \rightarrow bb$
cross section is approximately 
$\propto \lambda_{131}^{2}/ \lambda^{\prime 2}_{333}$. 
Therefore, even if one rescales down both
couplings (within a reasonable range), 
our results are only weakly affected.
It emphasizes the possible role of LEP if hints of an excess of neutral
current events at HERA are proved to be correct and are
interpreted within a supersymmetric framework~\cite{alt}.
(One may have to generalize the analysis to include $bd$ and $bs$ production.)

\subsection*{New fields}

Extensions of  $W_{\mbox{\tiny observable}}$ 
by new SM singlet and  vector-like fields 
at the weak or at intermediate scales
are also widely considered. A singlet superfield
could couple (via a Yukawa term) to $H_{1}H_{2}$ or to $LH_{2}$,
while new exotic charged fields could mix with the ordinary
SM fields. Their couplings could again affect the spectrum evolution.
(Exotic sectors could also contribute to the superoblique
parameters of Cheng {\em et al.}~\cite{so}.)
Additional weak-scale $U(1)^{\prime}$ models, which
could be motivated by string theory, often imply such new states
and were discussed in Ref.~\cite{esp}.
Here, however, we will discuss only the case of an intermediate-scale
right-handed (singlet) neutrino $N$:
$W_{\mbox{\tiny observable}} \rightarrow W_{\mbox{\tiny observable}}
+ M_{N}N^{2} - h_{N}H_{2}LN$.

As an example we have chosen
the possible interplay between the SSB parameters
$m_{N}^{2}$ and $m_{H_{2}}^{2}$.
While the SSB contribution to the intermediate-scale singlet
sneutrino mass is negligible, the size and sign of $m_{N}^{2}$ 
could be important for the evolution of the Higgs SSB parameters.
This is the case if $h_{N} = {\cal{O}}(1)$, as is suggested
by grand-unified models which typically predict $h_{N_{\tau}} = h_{t}$
at the grand-unification scale.
In our example, we  will further assume $m_{N}^{2} < 0$
and some (non-minimal) supergravity scenario
(and $|m_{N}^{2}| \sim {\cal{O}}(m_{3/2}^{2})$).

It was already shown in Ref.~\cite{FPT} that the evolution
of $m_{H_{2}}^{2}$ can be controlled and moderated by carefully
choosing $m_{Q_{3}}^{2}, m_{U_{3}}^{2} < 0$ at the Planckian boundary.
The physical stop squared masses are still positive
due to gluino loops (which are summed by the effective
potential or the renormalization), and  $m_{t}^{2}$ $F$-terms.
One replaces the tuning of electroweak parameters
({\em e.g.}, $\mu$, as discussed above)
with tuning of boundary conditions.
This can be understood intuitively: By allowing negative squared
masses at high scales, the $\Sigma_{m^{2}}$ terms in
eqs.~(\ref{rge1a}) -- (\ref{rge1c}) can be dialed to be
small, zero (which is in some sense a quasi-fixed point in the absence
of gaugino loops), or even negative. Due to the gaugino loops, $\Sigma_{m^{2}}$
can change sign in the course of integration.

Such dialing can be done more easily in the singlet
neutrino  framework ~\cite{FPT}.
Here, $m_{N}^{2}< 0$ could be a boundary condition or
could be achieved in the course of renormalization
between the Planckian and intermediate scales,
\begin{equation}
\frac{\partial m_{N}^{2}}{\partial \ln Q} = \frac{1}{8\pi^{2}}
(2h_{N}^{2}\Sigma_{m^{2}}^{N}),
\label{rge3a}
\end{equation}
where 
$\Sigma_{m^{2}}^{N} = [m_{H_{2}}^{2} + m_{L}^{2} + m_{N}^{2}+ A_{N}^{2}]$.
As noted above, $\Sigma_{m^{2}}$-type terms oscillate in sign
and hence the squared mass parameters are bounded from below.
For simplicity, let us assume that $m_{N}^{2} < 0$ is a boundary
condition. In either case, its effect on the physical heavy sneutrino
mass is negligible. We note in passing that if we did not introduce
the $M_{N}$ supersymmetric mass term, a radiatively generated
$m_{N}^{2} <0$ would lead to $\langle N \rangle \sim m_{3/2}$
and to new possibilities of neutrino mass generation~\cite{progress}.

The Higgs mass evolution is now given by 
\begin{eqnarray}
\frac{\partial m_{H_{2}}^{2}}{\partial \ln Q} = \frac{1}{8\pi^{2}}
(3h_{t}^{2}\Sigma_{m^{2}} + h_{N}^{2}\Sigma_{m^{2}}^{N}
& & \nonumber \\
- 3g_{2}^{2}M_{2}^{2} - g_{1}^{2}M_{1}^{2}). & &
\label{rge3b}
\end{eqnarray}
Clearly, by carefully choosing the boundary condition for $m_{N}^{2}$
(even for $m_{Q_{3}}^{2}, m_{U_{3}}^{2} > 0$)
one can control and moderate $m_{H_{2}}^{2}$ renormalization  at some
initial momentum interval, affecting the required tuning of
electroweak Higgs potential parameters.
Indirectly, $m_{N}^{2}< 0$ would also lead to lighter stops
(via eqs.~(\ref{rge1b}) and (\ref{rge1c})).
We conclude that the sneutrino presence is reflected
in this example in the magnitude  of the Higgs potential parameters.
Of course, $m_{L}^{2}$ renormalization is also affected.
However, we will return to this point after the following
and last example.

\subsection*{New sectors}

We briefly mentioned  above the possibility of new electroweak
and intermediate scale sectors. Here, however, we would
like to recall some of the consequences of a grand-unified
(GUT) sector with strong Yukawa interactions in supergravity 
models~\cite{oldies1,oldies2,PP}.
We will assume in this example supergravity and that the supergravity
scale ({\em i.e.}, the integration boundary) is above the GUT scale $M_{G}$.

One can distinguish two leading effects: Those which are due to large
SM Yukawa couplings, and in particular effects related to quark-lepton
unification, and those which are due to new large Yukawa couplings
which couple heavy states to light states.
The former is manifested in, ${\em e.g.}$,
\begin{equation}
\frac{\partial m_{10_{3}}^{2}}{\partial \ln Q}\left( Q > M_{G}\right) = 
\frac{1}{8\pi^{2}}(3h_{t}^{2}\Sigma_{m^{2}}^{10}  -...),
\label{rge4a}
\end{equation}
where we use  SU(5) classification, $10_{3} \ni (Q_{3},\, U_{3},\,
E_{3})$, $\Sigma_{m^{2}}^{10} = [m_{H_{2}}^{2} + 2m_{10}^{2} +
A_{t}^{2}]$, and we omitted the gauge terms. 
(In some extended models $L_{3}$ is also unified with the $t$-quarks
and ``$m_{L_{3}}^{2}(Q > M_{G})$'' is also renormalized
by $h_{t}^{2}$ terms.)
The effects are enhanced by the diverging $h_{t}(M_{G}) \sim 2$.
Eq.~(\ref{rge4a}) is valid between the GUT and supergravity (= string?) scales.
The latter effects are due to operators of the form $\lambda_{\Phi}\Phi H_{1}H_{2}$ 
where $\Phi$ is a heavy ${\cal{O}}(M_{G})$ field.
In minimal models at least one field $\langle \Phi \rangle \sim M_{G}$
and $\lambda_{\Phi} \sim 1$ are  required in order to render the color
triplet GUT-partners of the Higgs doublets sufficiently heavy
(so that radiative proton decay is sufficiently suppressed).
If indeed $\lambda_{\Phi} \sim 1$ then
\begin{eqnarray}
\frac{\partial m_{H_{2}}^{2}}{\partial \ln Q}\left( Q > M_{G}\right) 
=  \frac{1}{8\pi^{2}}
(C^{\Phi}\lambda_{\Phi}^{2}\Sigma_{m^{2}}^{\Phi} 
& & \nonumber \\ 
+ 3h_{t}\Sigma_{m^{2}}^{10}
- ...), & & 
\label{rge4b} 
\end{eqnarray}
where  $\Sigma_{m^{2}}^{\Phi} = [m_{H_{2}}^{2}+
m_{H_{1}}^{2}+m_{\Phi}^{2}+A_{\Phi}^{2}]$
and $C^{\Phi}$ is a group theory factor
(similarly for $m_{H_{1}}^{2}$).
Due to the large group dimensions and representations, all effects
are enhanced by large group-theory factors.
The combination of large couplings and large group theory factors
can more than compensate for the short integration interval.

It is interesting to note that the effects are somewhat complimentary:
$\lambda_{\Phi} \ll 1$ would suppress the second effect but
it leads to thresholds 
(corresponding to the GUT-partners of the Higgs doublets)
below the unification scale. In this case
there are threshold corrections $\propto h_{t}^{2}$
to the third family sfermion SSB mass parameters (particularly
to the slepton ones) which enhance the $h_{t}$ effects ~\cite{P}.
The threshold corrections are independent of the assumption
of renormalization beyond the GUT scale, which may not hold, for
example, in non-perturbative string theories.

It was recently realized that the heavy $t$-quark can lead to dramatic
effects in the renormalization of the third family sfermion 
and Higgs boson SSB parameters between the GUT and supergravity scales
(in particular, when combined with the effects of  large $\lambda_{\Phi}$
couplings). Even if one introduces the minimal supergravity 
universal boundary conditions at the supergravity scale, they are
wildly violated at the GUT scale due to their evolution
eqs.~(\ref{rge4a}) and (\ref{rge4b})~\cite{PP}.
In Ref.~\cite{PP} it was emphasized that the Higgs potential
parameters (and hence radiative symmetry breaking and 
fine-tuning considerations) as well as typical spectrum correlations
are significantly modified.
The latter may enable one to probe such scenarios. (See also Ref.~\cite{AN}.)
It is interesting to note that even if the $t$-quark were not heavy,
successful radiative symmetry breaking could be achieved in this and
in the previous  singlet neutrino example due to
$\lambda_{\Phi}\Sigma^{\Phi}_{m^{2}}$
($h_{N}\Sigma^{N}_{m^{2}}$) contributions to the evolution.

Ref.~\cite{BH}
focused on the corrections $\propto h_{t}^{2}$ to $m_{E_{3}}^{2}$,
which introduce flavor dependent corrections to the slepton 
spectrum. It was found that the corresponding contributions of slepton
loops to leptonic FCNC processes, particularly $\mu \rightarrow e \gamma$,
may not be negligible.
As we alluded to above, such corrections also arise in the case of
the singlet neutrino. Unlike the example here, the $h_{N}^{2}$
corrections split the left-handed slepton spectrum 
according to the hierarchies of  the singlet neutrino couplings and SSB 
parameters. (For a recent analysis, see Ref.~\cite{NNN}).
Potential contributions to flavor violation,
including $\mu \rightarrow e \gamma$~\cite{Hu}, also constrain
the couplings $\lambda$ and $\lambda^{\prime}$ 
of our first example (see above).
The couplings $\lambda^{\prime}$ also splits the left-handed
slepton spectrum, leading to a new class of contributions to FCNC
in lepton number violating models.
These contributions are not directly proportional to
the $\lambda^{\prime 2}$, are similar to the ones discussed in 
Ref.~\cite{BH,NNN}, and have not been studied yet.

\section{The Flavor Problem}

We concluded the previous section with the discussion of sparticle
loop contributions to leptonic FCNC processes. The contributions
discussed above were due to  radiatively generated 
non-universalities in the slepton spectrum\footnote{
Non-universalities $\propto
h_{t}^{2}\ln(M_{P}/m_{\mbox{\tiny{weak}}})$ 
in the squark spectrum can be typically ignored 
due to small 13 and 23 CKM mixing and hadronic
uncertainties. This is not the case here, given  the 
extremely sensitive search for, {\em e.g.},
$\mu \rightarrow e \gamma$.}.
Moreover, fermion and sfermion mass matrices
are not diagonalized simultaneously\footnote{The previous examples
implicitly assume non-diagonal lepton Yukawa matrices, as is suggested
by grand-unification. Any explicit predictions, {\em e.g.},
for $BR(\mu \rightarrow e \gamma)$ (as well as for radiative proton decay $p
\rightarrow K l$) must assume specific Yukawa textures.}
for generic non-universal (and non-proportional)
boundary conditions for the soft SSB parameters
$m_{i}^{2}$ (and $A_{ijk}$), leading to flavor non-diagonal
fermion-sfermion-gaugino (and sfermion-sfermion-Higgs) couplings.
In these cases sparticle loops can contribute 
to strongly constrained FCNC 
processes, in particular, those involving the first and second family fermions.

The unwanted contributions to FCNC processes
can be suppressed in two obvious ways:
$(a)$ Arranging for simultaneous diagonalization 
of fermion and sfermion mass matrices;
or  $(b)$ suppression of dangerous
loops in the case of arbitrary sfermion masses 
by raising the relevant sparticle mass scale
({\em e.g.}, $BR(\mu\rightarrow e \gamma) \propto 
1/m_{\mbox{\tiny{slepton}}}^{4}$).
In fact, these are the two mechanisms that are also invoked
to suppress dangerous contributions in the lepton number
violating models discussed above. 
However, there one can also suppress the new and arbitrary
Yukawa couplings $\lambda$ and $\lambda^{\prime}$,
while here the ordinary  Yukawa couplings are determined
by the fermion spectrum and are not arbitrary.
These solutions are perhaps the most important theoretical
criteria in constructing models for SSB mediation
to the observable sector. (See our discussion above
of minimal supergravity.) We will propose in the following section
new gravitational frameworks that can achieve these goals~\cite{NP}
at the minimal price of a low-energy singlet (rather than strong assumption on
the form of $K$).
Before doing so, we will briefly review some of the solutions,
their advantages, and their drawbacks.

\subsection*{Simultaneous diagonalization}

Two classes of sfermion mass matrices which commute with the Yukawa
matrices are often discussed - universal mass matrices (which are
proportional to the identity) and aligned mass matrices (sfermion 
mass matrices which are proportional to the Yukawa matrices).
Here, we will discuss only the former mechanism.

Imposing universality in the general supergravity framework
leads to the minimal framework described above.
It requires strong assumptions on the structure of the Kahler
potential, and in particular, suppression of any light-heavy
mixing which generically leads to $\Lambda_{ab} \neq \delta_{ab}$~\cite{SW}.
It was recently pointed out that even if such assumptions are made,
gravitational interactions typically generate radiatively flavor
non-diagonal masses $(\Delta m^{2}/m^{2}) \sim (N/16\pi^{2})$,
unless the model has a stringy $T$-duality~\cite{NP}. 
(Here $N = {\cal{O}}(10)$ counts the number of massless states.)
These radiative corrections are gravitational
and are different from the model-dependent field-theory radiative corrections
that we discussed in previous examples and which are still in effect.
The dilaton limit in perturbative string theory
corresponds to a (calculable) example of a universal model~\cite{jan1}.
It is also subject to (stringy) radiative corrections~\cite{jan2}.
In addition, no convincing realization of this limit in string theory exists.
More generically, perturbative string theory predicts a 
non-universal spectrum which depends on the modular weights of the fields.
(This seems also to be the situation in non-perturbative string
theories~\cite{M}.) In the {\em a priori} 
unlikely case of universal weights one
would recover the supergravity limit $\Lambda_{ab} = \delta_{ab}$.

It is sufficient, however, to require only charged-dependent 
flavor universality. In (any) supergravity framework
this is the case if $M_{1/2} \gg m_{i}(0)$ where $m_{i}(0)$ is a generic 
boundary condition for the soft SSB scalar masses.
In string theory this would be the case
if there is a symmetry that relates the gauge charges and the modular 
weights~\cite{NP}. This is also the situation
in the framework of low-energy supersymmetry breaking
if SSB is mediated to a messenger sector ({\em i.e.}, 
$W \neq W_{\mbox{\tiny hidden}} \oplus W_{\mbox{\tiny observable}}$)  
and then to the observable sector via gauge interactions
({\em i.e.}, gauge mediation of SSB).
(For reviews and references, see Ref.~\cite{GM,Dine}.) In fact, only the nature
of  the second mediation is relevant for the issue of universality,
a point which is relevant to  the discussion in the next section.
In all such examples one expects mass hierarchy according
to charges, {\em e.g.}, heavy squarks and significantly lighter sleptons.

Some elaboration on the 
the low-energy messenger/gauge-mediation framework, 
which offers an alternative
phenomenological framework to supergravity models, is in place.
The gaugino masses result in this framework
from the messenger (SM) gauge interactions and are generated  at one-loop,
$m_{\mbox{\tiny{gaugino}}}(\Lambda) \sim (g^{2}/16\pi^{2})\Lambda$.
The scalar squared masses, on the other hand, are generated only at two-loops,
$m_{i}^{2}(\Lambda) \sim [(g^{2}/16\pi^{2})\Lambda]^{2}$ 
(since one typically forbids
direct interactions between the messengers and the ordinary matter fields
which could be flavor dependent).
Therefore, gaugino masses and scalar masses are
of the same order of magnitude without any special assumptions
on the gauge kinetic functions $f^{i}_{\alpha\beta}$. 
(This is also the case in the dilaton limit~\cite{jan1}
discussed above and, apparently, in non-perturbative string theory~\cite{M}.)
The scale $\Lambda$ typically corresponds to the SSB mediation scale,
which is roughly of the order of the SSB scale.
One finds that such models are phenomenologically valid
if $\Lambda \sim 10^{5}$ GeV, which is well below the unification
and intermediate scales. 
(The heavy squarks compensate in this 
case for the shorter integration interval in the radiative symmetry
breaking mechanism.)
On the other hand, in the low energy SSB case, 
$m_{3/2} \sim W_{\mbox{\tiny hidden}}/M_{P}^{2} \sim \Lambda^{3}/M_{P}^{2}
\rightarrow 0$. 
As a result, supergravity mechanisms to generate
$\mu \sim m_{3/2}$ are irrelevant (like most other supergravity
contributions). Radiative generation of $\mu$ at low energy
typically implies the generation of ``$B\mu$'' at the same loop order
and $``B\mu$''$/\mu \sim \Lambda$. Such a situation leads to
an unacceptable fine-tuning. This is perhaps the most severe
phenomenological difficulty in this framework~\cite{Dine}.
We will return to the gauge mediation/messenger framework
in the next section.

\subsection*{Arbitrary sfermion mass matrices}

The phenomenological ``2--1'' framework that underlies these ideas is
straightforward and was reviewed in Ref.~\cite{Nelson}.  One
distinguishes a heavy decoupled sector which contains the first and
second family sfermions (``2'' generations) and a lighter sector which
contains the Higgs fields, as well as the gauginos and third family
(``1'' generation) fields which couple to the Higgs fields (see
eq.~(\ref{rge1a})).  The most dangerous sparticle contributions to
FCNC processes (and to CP odd observables) decouple, while {\em a
priori} no extreme fine-tuning is required in the Higgs potential.
The realization of such a framework, however, is not as
straightforward.  Such models could be realized in the framework of
dynamical low-energy supersymmetry breaking~\cite{Nelson} or by
embedding an anomalous U(1) horizontal symmetry in a supergravity
framework~\cite{DPDP}. A new realization of this framework will be
discussed in the next section.

The two sectors, of course,  are not truly decoupled. Terms which couple
their renormalization and are typically suppressed by small 
hypercharge and Yukawa couplings, two-loop factors
and by the universality assumption, are now enhanced 
by the heavy masses and their arbitrariness.
Hence, in practice there is a trade off between fine-tuning and
an upper bound on the decoupled sector scale 
(and hence, the spectrum  arbitrariness)~\cite{DGDG}.
(The bound  depends on the specific realization and on the
relevant integration  interval.) It appears that
more severe constraints on the decoupling scale
are due to negative physical stop squared masses
~\cite{AM}. However, note that in the derivation of the latter constraints,
Yukawa terms were omitted even though negative squared-mass parameters
appear (either radiatively or as boundary conditions). 
Therefore, such omission is not justified~\cite{KP}.
(See an example in the previous section).

We note in passing that
this and some of the previous frameworks contain heavy states
that may be probed by the superoblique parameters. The implications 
of the flavor problem and its solutions to the dimensionless
couplings and to the superoblique parameters are discussed 
in Ref.~\cite{so}.

\section{Gravitational triggering models}

Solutions to the flavor problem, which we discussed above,
often assume contributions to the scalar spectrum which are
independent of the Kahler potential, so that no strong
assumptions on the form of $K$ need to be made.
(But see our proposal of a symmetry relating modular weights
and gauge charges in string theory. In general, some symmetry
assumptions have to be made somewhere.)
The gauge mediation (messenger) mechanism and some 
realizations of the decoupling
mechanism further assume that gravitational interactions
have no significant role in the mediation of SSB.
That is, it is usually assumed that gravitational interactions are negligible
for the determination of the spectrum if supersymmetry is broken
at low-energies.
This intuitive assumption holds in general, but it fails
in the particular case of a light global and local (universal) singlet.

Here, we would like to show in a simple example how
such a singlet can lead to gravitational triggering of
a gauge mediation framework. We will apply the general framework
to two specific examples: An ordinary messenger framework
and  a horizontal messenger framework. Our discussion follows
the work of Ref.~\cite{NP}. 
(Ref.~\cite{NP} also contains an application
which we do not discuss here and which 
attempts to solve the ``low-energy $\mu$ problem''
described in the previous section.)

The observable sector is extended to include an  universal singlet,
$S = s +\theta^{2}F_{s}$, and a  vector-like pair of non-singlet fields
$V$ and $\bar{V}$.
The specific toy model that we will consider here is given
by the superpotential
\begin{equation}
W_{\mbox{\tiny observable}} \rightarrow
W_{\mbox{\tiny observable}} + \frac{\kappa}{3}S^{3} +
\lambda_{S}SV\bar{V}. 
\label{W}
\end{equation}
The Kahler potential is given by (suppressing
higher orders in $M_{P}^{-1}$)
\begin{equation}
K = \sum_{I}
\left(1 + \alpha_{I}\frac{S + S^{\dagger}}{M_{P}} \right)
\Phi_{I}\Phi^{\dagger}_{I}.
\label{K}
\end{equation}
Obviously, we assume no ${\cal{O}}(M_{P})$ dimensionful parameters for
the singlet field, a situation which can be understood in the
context of a string theory or duality transformations.
The $D$-terms
\begin{equation}
\int d^{2}\theta d^{2}\bar{\theta}e^{K/M_{P}^{2}}K
\label{D}
\end{equation}
lead to a quadratically divergent (tadpole) contribution to
the scalar potential~\cite{BPR},
\begin{equation}
\Delta V_{\mbox{\tiny{SSB}}} = -\frac{M_{SUSY}^{4}}{M_{P}}s + h.c.,
\label{tadpole}
\end{equation}
where we implicitly assume
no Planckian values for suspersymmetry breaking fields.
(The general situation, as well as  our phase choice
and the ambiguity due to dimensionless couplings and loop factors, 
are discussed in Ref.~\cite{NP}.)
Including the usual $F$-terms $|\partial W/ \partial S|^{2} =
\kappa^{2}s^{4}$ 
(which is the only relevant term if $2\lambda_{S} > \kappa$) and neglecting 
any other supergravity terms aside from the tadpole,
we have 
\begin{equation}
s = \left(\frac{M_{SUSY}^{4}}{4\kappa^{2}M_{P}}\right)^{\frac{1}{3}}, \,\,
F_{s} = \kappa  s^{2},
\label{vev}
\end{equation}
and the scalar messenger squared-mass matrix
\begin{equation}
M^{2}_{v\bar{v}} \sim 
\left(\begin{tabular}{c c}
$\lambda_{S}^{2}s^{2}$& $\lambda_{S} F_{s}$\\
$\lambda_{S} F_{s}^{*}$ & $\lambda_{S}^{2}s^{2}$
\end{tabular}\right).
\label{Mv}
\end{equation} 
The diagonal term is a supersymmetric mass term,
{\em i.e.}, the corresponding fermions have a similar Dirac mass term
$\lambda_{S} s \tilde{v}\tilde{\bar{v}}$.
Similarly, field dependent masses are induced for $s$ and for its fermion
partner $\tilde{s}$, and are given by replacing $\lambda_{S}$ with
$\kappa/3$ ($\kappa$) in the diagonal (off-diagonal) 
terms in eq.~(\ref{Mv}).

The fields $S$ and $V,\,\bar{V}$
are suitable to play the role of the singlet and non-singlet messenger
fields in a gauge mediation framework with (the usual ~\cite{GM})
$\Lambda = F_{s}/s \sim s$.
Note the minimality of this framework 
as SSB is transmitted to the messenger fields
gravitationally, {\em i.e.},
$W = W_{\mbox{\tiny hidden}} \oplus W_{\mbox{\tiny observable}}$.
Only the fields $S,\, V,\,\bar{V}$ are needed and
the messenger sector is absorbed in the observable sector.
The quadratic divergences allow for the gravitational mediation of
(low-energy) SSB to the messenger singlet, hence triggering
the usual gauge-mediation framework.

\subsection*{ A messenger model}

By examination of the gravitationally triggered gauge mediation
framework (\ref{vev}) -- (\ref{Mv}),
we find that $M_{SUSY} \sim 10^{8}$ GeV
corresponds to $\Lambda \sim 10^{5}$ GeV.
($M_{SUSY} \sim 10^{5}$ GeV, had we allowed Planckian values
for supersymmetry breaking fields.)
Hence, we find the usual  messenger model framework~\cite{GM,Dine}.
It depends only on the scale of SSB, not on its details.

\subsection*{A horizontal messenger model}

If  $M_{SUSY} \sim 10^{10}$ GeV,
supergravity masses are not negligible.
However, from eq.~(\ref{vev}) we have in this case $\Lambda \sim 10^{7}$ GeV,
which leads to ${\cal{O}}(10$ TeV) masses in the corresponding
gauge mediation framework. (Corrections $\sim m_{3/2}$
to (\ref{vev}) are still negligible.)
This numerology hints at a new realization
of decoupled $``2-1$'' models.
We can assume that $V$ and $\bar{V}$ are SM singlets
but are charged under some horizontal symmetry
that distinguishes the first two families
(which transform non-trivially) from the third family and Higgs fields
(which transform trivially). In this case, the former
would acquire ${\cal{O}}(10$ TeV) masses via 
(horizontal) messenger gauge loops while the latter would have only 
${\cal{O}}(m_{\mbox{\tiny{weak}}})$
supergravity masses.
(In the case of Planckian values for supersymmetry breaking
fields, $\Lambda \sim M_{SUSY} \sim 10^{10}$ GeV
can be offset by an extremely weak horizontal gauge coupling.)

\section{Summary}

In summary, we discussed some issues
in the evolution and sources of the superparticle
spectrum. The flavor problem was shown to be affected by the 
former and to be a useful criterion in classifying
and selecting the latter.
Various unrelated effects, from light sneutrinos
to radiative contributions to FCNC, have been shown to
originate from similar mechanisms.
New sources of flavor violation were identified.
The flavor problem  (which is often attributed
to the {\em ad hoc} arbitrariness of the Kahler potential)
was discussed in some detail,
and it was shown that some of its solutions can be triggered
gravitationally (from the divergent 
Kahler potential interactions of a universal singlet).

Underlying our discussion and the various examples are
the potential difficulties in the theoretical interpretations
of future measurements. Indeed, each interpretation requires
one to specify a framework for the renormalization of the spectrum
({\i.e.}, the non-hidden sector)
and a framework to the mediation of SSB to the observable sector
({\em i.e.}, the boundary and the 
parameterization of the  boundary conditions).

In the previous section we considered 
examples of gravitationally triggered frameworks which  suggest
that the supergravity, $M_{SUSY}\sim 10^{10-11}$ GeV, and the
low-energy SSB framework, $M_{SUSY}\sim 10^{5-7}$ GeV, only mark the
most simple (and perhaps most attractive) limits on a scale chart.
(See Ref.~\cite{others} for different examples.)  In the simplest
versions of these limits the spectrum is described by a small number
of parameters, ($m_{0},\, A_{0},\, B, M_{1/2},\, \mu$) and
($\Lambda,\, B,\,\mu$), respectively.  The horizontal messenger
example suggests a linear combination of (new) gauge mediated and
(non-minimal) supergravity contributions to the spectrum, and some
states beyond the reach of future colliders.  The parameterization 
of the spectrum may
not be as simple in this case, and some of the parameters may never be
characterized by experiment (but see Ref.~\cite{so}).  Even if the
simple low energy SSB + gauge mediation framework describes the data
in a satisfactory fashion, our messenger example suggests that its
interpretation is not unique.

In addition, the need to specify a renormalization framework
(one usually assumes the minimal supersymmetric SM framework)
can lead to new difficulties. On the other hand,
it provides a rare opportunity to learn about the non-hidden sectors
at intermediate and high energies.
Hence, our window on the observable/non-hidden sector
comes at the price of a potentially 
less ``clear view'' of the SSB/hidden sector.
The correct renormalization framework 
may reveal itself in the identity of the (N)LSP, as well as
in the enhancement of rare processes such as $\mu \rightarrow e \gamma$
(though in both examples the observations
may not have a unique interpretation).

It is important to keep in mind
that analysis within the most simple frameworks, though it is a
well motivated first step, is far from unique and it may 
partially or completely  fail. 
Such failure may be due to simplistic assumptions,
and alternative interpretations should then be pursued.
Particularly, we argued that the LEP collaborations
should exploit their physics potential and 
search for a singly produced sneutrino if conventional sparticle searches fail.

The examples explored in this contribution  are based on old and new works in
various collaborations with 
H.-C.~Cheng, J.~Erler, J.~L.~Feng, C.~Kolda, H.~P.~Nilles,
A.~Pomarol, and S.~Thomas.

\end{document}